\documentstyle[preprint,prl,aps]{revtex}
\begin{document}

\title{Current-Loop Model for the Intermediate State \\ of Type-I
Superconductors}

\author{Raymond~E. Goldstein$^{1}$, David P. Jackson$^{1*}$, and
Alan T. Dorsey$^{2}$}

\address{$^{1}$Department of Physics, Joseph Henry Laboratories,
Princeton University, Princeton, NJ 08544}

\address{$^{2}$Department of Physics,  University of Virginia, McCormick Road,
Charlottesville, VA 22901}

\date{\today}

\maketitle

\begin{abstract}

A theory is developed of the intricately fingered
patterns of flux domains observed in the intermediate state of
thin type-I superconductors.  The patterns are shown to arise
from the competition between the long-range Biot-Savart interactions of the
Meissner currents encircling each region and the superconducting-normal
surface energy.  The energy of a set of such
domains is expressed as a nonlocal functional of the positions of their
boundaries, and a simple gradient flow in configuration
space yields branched flux domains qualitatively like those seen in experiment.
Connections with pattern formation in amphiphilic monolayers and
magnetic fluids are emphasized.

\end{abstract}

\pacs{PACS numbers: 05.70.Ln, 74.55.+h, 75.60.-d}

When a thin film of a type-I superconductor is placed in a magnetic field
normal to the sample, the large demagnetizing effects associated with the
film geometry preclude the establishment of the Meissner phase (with
magnetic induction ${\bf B}=0$).
The sample instead accommodates the field by breaking up into a large number of
superconducting (${\bf B}=0$) and normal (${\bf B}\neq 0$) regions,
usually forming very intricate patterns \cite{Huebener}.
These arise in part from the
competition between the magnetic field energy of the domains and the
surface energy between the superconducting and normal regions.
However, they have not been successfully
explained theoretically, for virtually all theories of these
patterns, starting from that of Landau \cite{Landau},
have explored this competition with variational calculations that
assume {\it regular} geometries
of the flux domains \cite{others}.
The hypothesized parallel stripes, ordered arrays of circles, etc. are rarely
seen, the norm instead being the disordered patterns documented in classical
experiments \cite{Huebener,Haenssler,Faber}.
Moreover, the temperature and magnetic field history of the sample
strongly influence the patterns, suggesting that
they are likely not in a global energetic minimum.

Recent work in this area has emphasized that the diffusion
of magnetic flux in the normal phase can influence the domain morphology
\cite{Dorsey,Goldenfeld}.
{}From asymptotic methods applied to the time-dependent
Ginzburg-Landau (TDGL) model \cite{asymptotics1,asymptotics2},
a free-boundary representation for the motion of superconductor-normal
(S-N) interfaces has been derived which
is nearly identical to those describing the growth of a solid
into a supercooled liquid, where the interface motion is
unstable (producing dendrites, for example).  By analogy it was
suggested, and confirmed by numerical solution
of the TDGL equations \cite{Dorsey,Goldenfeld},
that the growth of the superconducting phase into the supercooled
normal phase should be dynamically unstable, leading to highly ramified
domain shapes.  While such diffusive
instabilities may play a role in the pattern formation in the intermediate
state, these studies are not directly applicable as they have all
ignored demagnetizing effects.

Here we take a different approach to
the intermediate state, ignoring entirely the diffusion
of flux and focusing instead on the demagnetizing fields.
We address the basic question: {\it What is the energy of a thin
multiply-connected superconducting domain, the normal regions of which
are threaded with a magnetic field?}
A central issue is the degree to which
the interactions between the Meissner currents, flowing along the
S-N interfaces within the film, are screened by the
superconducting regions.  Pearl \cite{Pearl} and Fetter and Hohenberg
\cite{FetterHohenberg} made the important observation that vortices in a thin
film interact with a
potential $V(r)\sim 1/r$ for large separations $r$, while for small $r$,
$V(r)\sim \ln (\Lambda/r)$, with $\Lambda$ an appropriate cut-off.
Unlike in bulk, the interactions are unscreened,
being dominated by the electromagnetic fields
above and below the film.  This
suggests the simple model developed here:
domains bounded by current loops interacting as in free space, endowed with
line tension, and subject to the constraint of constant total magnetic flux
through the sample.  We focus on how such loops
relax toward a local energy minimum when prepared in a nonequilibrium
configuration.  Such a model allows for the
study of the shape of the domains,  but does not readily account for changes in
topology (e.g., fission).  We conclude that the long-range
interactions destabilize flux domains of regular shape,
producing branched, fingered structures as seen in experiment.

Apart from the global flux constraint,
this model is equivalent to one for domains of
magnetic fluids \cite{Rosensweig} in Hele-Shaw flow \cite{cebers,pra},
where experiments \cite{ferroexpt1,ferroexpt2} show patterns
like those observed in the intermediate
state, with the same history-dependence noted earlier.
Thin magnetic films
\cite{Seulfilms} and monolayers of
dipolar molecules \cite{monolayers} exhibit similar behavior, and are described
by such models through the underlying similarity between electric and magnetic
dipolar
phenomena \cite{Andelman,flucpap}.  This model is also very similar
to one introduced for pattern formation in certain reaction-diffusion
systems \cite{turingpap}.  There, fronts between regions of different chemical
composition move in response to line tension and a nonlocal Biot-Savart
coupling, and may exhibit labyrinthine patterns, as
observed in recent experiments \cite{Lee}.

To begin, we note a crucial separation of length scales
between the typical size of the flux domains (ca. $0.1$ mm)
and the penetration depth (ca. $\le 1$ $\mu m$)
\cite{Huebener}.  Thus {\it on the scale of the patterns the
superconductor-normal interface is sharp, and we may view the order
parameter magnitude as piecewise constant}. It follows that the
energy is determined solely by the locations of the S-N interfaces.

In this macroscopic approach, the
energy ${\cal F}$ of a configuration of flux domains
arises from the bulk free energy
of the two coexisting phases, the boundaries
between them, and the magnetic field energy,
\begin{equation}
{\cal F}={\cal F}_{\rm bulk}+{\cal F}_{\rm int}+{\cal F}_{\rm field}~.
\label{total_energy}
\end{equation}
Suppose the film has thickness $d$, total area $A$, volume $V=Ad$,
and contains a set of normal
domains $i$ with area $A_i$, length
$L_i$, and whose boundary positions are
${\bf r}_{i}(s)$ (Fig. \ref{model_fig}).  We assume ${\bf r}_{i}$ is
independent of $z$, neglecting
the ``fanning out'' of the domains near the film surfaces
\cite{Landau}.
The two phases occupy volumes $V_s$ and $V_n=d\sum_iA_i$,
with $V_s+V_n=V$.  Their
bulk free energy densities $F_s$ and $F_n$ define the critical
field $H_{c}(T)$ as
$F_n-F_s=H_c^2/8\pi$.  With
$\rho_n=A_n/A$ the area fraction
of the normal phase, $\sigma_{\rm SN}=(H_{c}^{2}/8\pi)\Delta$ the S-N
interfacial
tension (where $\Delta(T)$ can be interpreted as the interfacial width), we
have
\begin{equation}
{\cal F}_{\rm bulk}=\left[F_s+{H_c^2\over 8\pi}\rho_n\right]Ad~;
\ \ \ {\cal F}_{\rm int}= {H_{c}^{2} d\over 8\pi} \Delta \sum_i L_i~.
\label{bulk_energy}
\end{equation}

The complexity of this problem lies entirely in the computation of the
field energy.
An applied field ${\bf H}=H_a {\bf \hat e}_z$,
produces a field $H_n{\bf \hat e}_z$ in the normal regions, where
$AH_a=A_nH_n$ by flux conservation, so
\begin{equation}
H_n={H_a\over \rho_n}~.
\label{flux_conservation}
\end{equation}
The requirement of tangential continuity of ${\bf H}$ across
a S-N interface gives the field in the superconducting region
as ${\bf H}_{s}= H_s {\bf \hat e}_z=H_n {\bf \hat e}_z$.  Taking the
superconducting regions to be perfectly diamagnetic (with susceptibility
$\chi=-1/4\pi$) yields a magnetization ${\bf M} = -(H_{n}/4\pi) {\bf \hat
e}_z$,
which guarantees that the magnetic induction in the
superconducting regions
vanishes: ${\bf B}_s={\bf H}_s+4\pi {\bf M}=0$.
The field energy is that of
domains of total magnetization ${\bf M}_{\rm tot}$ (as yet unknown)
in an external field ${\bf H}_{a}$,
\begin{equation}
{\cal F}_{\rm field}= -{1\over 2}\int\! d^3r {\bf M}_{\rm tot}\cdot {\bf
H}_{a}~.
\label{magenergy}
\end{equation}

We quickly point out that if we ignore the fringing
fields entirely and assume ${\bf M}_{\rm tot}$ is completely uniform,
the field energy would be
\begin{equation}
{\cal F}_{\rm field}'= -{1\over 2} V_s \left( - {H_{n}\over 4\pi}\right) H_{a}
              = {H_a^2\over 8\pi}{1-\rho_n\over\rho_n} Ad.
\label{nodemag}
\end{equation}
Minimizing the sum ${\cal F}_{\rm bulk}+{\cal F}_{\rm field}'$
sum with respect to $\rho_{n}$, we obtain the standard result \cite{Tinkham}
for the equilibrium value $\rho_{n}^{*} = H_{a}/H_{c}$;
i.e., $H_{n} = H_{c}$.  But we obtain no information about the shapes of
the domains.  The inclusion of line tension alters slightly the
equilibrium area fraction, and renders the equilibrium domain shape a
circle, as that has the minimum perimeter for a given area.

A field energy sensitive to the shape of the
flux domains must account for fringing magnetic fields.
Since the calculation of these is well-documented \cite{cebers,pra},
we summarize only the essential results.
First write the total magnetization as ${\bf M}_{\rm tot}=\chi{\bf H}_{\rm
cap}$,
where ${\bf H}_{\rm cap}$ is the magnetic field arising from the
uniform magnetization
$\chi {\bf H}_n$. The subscript ``cap'' indicates the equivalence
of this problem to that of a uniformly charged parallel plate capacitor.
Now write ${\bf H}_{\rm cap} ={\bf H}^{(0)}+{\bf H}^{(1)}$, where
${\bf H}^{(0)}=H^{(0)}{\bf \hat e}_z$ is uniform between the plates and
vanishes elsewhere.  The choice
$H^{(0)}=-4\pi\chi H_n$ yields $\bbox{\nabla}\cdot{\bf H}^{(1)}=0$,
so ${\bf H}^{(1)}=\bbox{\nabla}\times {\bf A}$, for some vector
potential ${\bf A}$. Choosing an appropriate gauge, we find
$\nabla^2 {\bf A}= \bbox{\nabla}\times {\bf H}^{(0)}$, or
\begin{equation}
{\bf A}({\bf r}) = -{H_{n}\over 4\pi} \sum_i\int_0^d dz\oint \!
ds{{\bf \hat t}_i(s)\over \vert {\bf r}-{\bf r}_i(s,z)\vert}~,
\label{vector_potential}
\end{equation}
where ${\bf \hat t}_i(s)$ is the tangent to ${\bf r}_i(s)$.
Substituting into (\ref{magenergy}), performing the integrations over the
thickness
of the slab \cite{pra}, the energy of a set of
interacting current loops is found to be
\begin{eqnarray}
{\cal F}[\{{\bf r}_i\}] &=& {H_c^2 d\over 8\pi}\Biggl\{\Delta\sum_i L_i+
A\left[\rho_{n} +{h^2\over \rho_n}\right]
\nonumber \\
&&\qquad -{h^2\over 4\pi\rho_{n}}\sum_{i,j}
\oint\!ds\oint\!ds'\,{\bf \hat t}_i\cdot{\bf \hat t}_j\, \Phi_{ij}\Biggr\}~,
\label{totalenergy}
\end{eqnarray}
where $h=H_a/H_c$ is the reduced applied field,
${\bf \hat t}_i={\bf \hat t}_i(s)$,
and $\Phi_{ij}=(1+\delta_{ij})\Phi(R_{ij}/d)$, with
$R_{ij}=\vert {\bf r}_i(s)-{\bf r}_j(s')\vert$.  The interaction potential
\begin{equation}
\Phi(\xi) = \sinh^{-1}\left(1/\xi\right) + \xi - \sqrt{1+\xi^2},
\label{Phi}
\end{equation}
is Coulombic for $\xi\gg 1$, $\Phi\approx 1/(2\xi)$,
reflecting the usual self-induction energy for elementary currents,
while for $\xi \ll 1$,  $\Phi\approx \ln (2 e^{-1}/\xi)$, with the
film thickness $h$ acting as a cutoff
(as in the vortex potential $V(r)$ above).

The motion of S-N interfaces is determined by the normal component of
the force acting
at a point ${\bf r}_i(s)$ on a given interface.  It is
the negative functional derivative of the energy \cite{asymptotics2,pra}
\begin{eqnarray}
 - {1\over \sqrt{g_{i}}}{\bf \hat n}_i\cdot {\delta {\cal F} \over \delta {\bf
r}_{i}}
& = & {H_c^2d\over 8\pi}\Biggl\{-\Delta{\cal K}_{i}(s)+\Pi \nonumber \\
&& -{h^2\over 2\pi\rho_n d}\sum_j \oint\!ds'_j\,
{\bf \hat R}_{ij} \times{\bf \hat t}_{j} \Phi_{ij}'\Biggr\}~,
\label{U0}
\end{eqnarray}
where $g_{i}=\partial_{\alpha}{\bf r}_{i}\cdot\partial_{\alpha}{\bf r}_{i}$
is the metric, ${\cal K}_{i}(s)$ the curvature, ${\bf \hat n}_{i}$ the unit
normal,
${\bf \hat R}_{ij}={\bf R}_{ij}/\vert {\bf R}_{ij}\vert$,
$\Phi'(\xi)=
1-(1+\xi^{-2})^{1/2}$, and
$\Pi$ is an effective pressure given by
\begin{equation}
\Pi = {h^{2}\over \rho_{n}^{2}}\left[1-
{1\over 4\pi A} \sum_{i,j}\oint\!ds\oint\!ds'\,
{\bf \hat t}_i\cdot{\bf \hat t}_j\, \Phi_{ij}\right]-1~.
\label{pressure}
\end{equation}

Equation (\ref{U0}) reveals that the pattern formation arises from
a competition between a magnetic pressure which incorporates flux
conservation, the
Young-Laplace force due to interfacial tension
and the Biot-Savart force from the circulating currents.  This additional
long-range contribution to the boundary force has been noted before
for magnetic fluids \cite{cebers,pra}
and amphiphilic monolayers \cite{Kessler}.
{}From extensive studies \cite{cebers,pra,ferroexpt1,ferroexpt2,flucpap}
of this competition in domains of {\it fixed} area, we know of
the existence of branching instabilities of circular domains and buckling
instabilities of stripes, phenomena which should carry over to the
present problem with nonconserved area. Indeed, buckled domains are well known
in type-I superconductors \cite{Haenssler,Faber} and are seen as well in
amphiphilic monolayers \cite{Stine}.

The simplest dynamical model for interface motion
balances the normal component of the force against a viscous force
localized on the interface, yielding an equation of motion \cite{pra}
\begin{equation}
\eta\, {\bf \hat n}_{i} \cdot {\partial {\bf r}_i(s)\over \partial t} =
- {1\over \sqrt{g_{i}}}{\bf \hat n}_{i}
\cdot {\delta {\cal F} \over \delta {\bf r}_{i}}~,
\label{eom}
\end{equation}
where $\eta$ is a kinetic coefficient.
This equation of motion is local in time; in the present context it arises
from the neglect of the diffusion of the magnetic flux in the normal
phase \cite{asymptotics2},
which is equivalent to assuming that the normal state
conductivity is zero.  The magnetic vector potential is then
``slaved'' to the order parameter, a limit
considered also in the reaction-diffusion context\cite{turingpap}.
To estimate $\eta$, we appeal to earlier results in the slaving
limit for strongly type-I {\it bulk} superconductors \cite{asymptotics2}
and obtain
\begin{equation}
\eta = {H_c^2d \Delta\over 8\pi}{\pi \hbar\over 8 k_BT_c \xi_0^2}~,
\label{eta}
\end{equation}
where $T_c$ is the critical temperature and $\xi_0$ is the zero-temperature
correlation length \cite{abrikosovbook}.
These results can be used to estimate the time scales for domain motion
from other experimentally measured parameters.

The dynamical evolution of flux domain shapes
governed by Eqs. (\ref{U0})-(\ref{eom})  is a many-body problem of considerable
computational complexity.  To gain insight into its behavior, we consider
here a mean-field description in which only the self-interaction of the
current loops is considered;  the Biot-Savart interactions between loops are
neglected, while the flux-conservation condition is enforced by
assigning the loop to a cell of area $A_{\rm cell}$,
with $\rho_n=A_n/A_{\rm cell}$.  By appropriate rescaling of the
spatial variables,
we deduce from Eqs. (\ref{U0}) and (\ref{pressure}) that the location
of the system in the $H-T$ plane is uniquely specified by the two
dimensionless quantities $h$ and $\Delta/A_{\rm cell}^{1/2}$.

To see the effects of the Biot-Savart interaction on the stability of
a circular flux domain, Figures \ref{sim_fig}a and b show the evolution of a
domain prepared with an area significantly less than the equilibrium
value.  In Fig. \ref{sim_fig}a the Biot-Savart coupling is omitted,
and the circle simply relaxes to a new radius driven primarily by the magnetic
pressure.  Including the long-range interactions (Fig. \ref{sim_fig}b)
results instead in  the formation of a branched flux
domain with three-fold coordinated nodes, very similar to those seen in
experiment \cite{Huebener,Haenssler,Faber}.
In the early epoch the shape evolution is primarily a dilation with
little change in shape, while the branching instabilities occur on a
longer time scale.
As found in previous studies \cite{ferroexpt2},
in the later stages of the shape evolution the driving force for the
interfacial motion becomes extremely small, with very small energy
differences between rather different shapes.  This
suggests that the interface motion would be extremely sensitive
to external perturbations such as impurities or grain boundaries, which
would then be effective in pinning the interfaces, not unlike the pinning of
vortices in type-II superconductors \cite{phil}.  This may contribute to the
history-dependence of the patterns discussed in the introduction.

Elsewhere we will discuss
the derivation of the present model from the
more fundamental TDGL equations by generalizations
of existing asymptotic methods \cite{asymptotics1,asymptotics2,turingpap},
as well as the details of stability analyses for various regular
geometries.   Given the connections outlined here between pattern formation
in the intermediate state and in other systems such as amphiphilic monolayers
and magnetic fluids, we suggest that it is of interest to extend
experimental studies of flux domain shapes to probe systematically
\cite{Seulfilms} the
branching instabilities as a function of applied field and
temperature.

We are grateful to T.C. Lubensky, N.P. Ong, and V.S. Pande for important
comments at an early stage of this work, and thank A.O. Cebers, S. Erramilli,
D.M. Petrich, M. Seul, M.J. Shelley, and T.A. Witten for numerous discussions.
This work
was supported in part by NSF Presidential Faculty Fellowship
DMR 93-50227 (REG),  NSF Grant DMR 92-23586 (ATD), and Alfred P. Sloan
Foundation Fellowships (REG and ATD).

\begin{figure}
\caption{A thin slab of a type-I superconductor, of thickness $d$,
viewed along the applied field ${\bf H}_{a}$.  The normal regions are
shown shaded.  Adapted from \protect{\cite{Haenssler}}.\label{model_fig}}
\end{figure}

\begin{figure}
\caption{Numerical results from the gradient-flow model for
the evolution of a flux domain boundary. Dashed initial condition
is a circle of unit radius perturbed by low-order modes, evolving
with $\Delta=0.01$ and $h=0.5$.  (a)  Rapid relaxation to a circle in the
absence of long-range forces. (b) Rapid dilation followed by
fingering on a much longer time scale in the presence of
the Biot-Savart interactions.
\label{sim_fig}}
\end{figure}

\end{document}